\documentclass{evn2024}
\usepackage{graphicx}
\usepackage{epstopdf}
\begin{document}
   \title{Investigating launching of black hole jets with the combined power of the EVN and the EHT}
    \titlerunning{Jets seen with EHT \& EVN}
   \author{G.~F. Paraschos\inst{1}
          \and
          L.~C. Debbrecht\inst{1}
          \and
          J.~A. Kramer\inst{2,1}
          \and
          E. Traianou\inst{3,1}
          \and
          I. Liodakis\inst{4,5}
          \and
          T. Krichbaum\inst{1}
          \and
          J.-Y. Kim\inst{6,1}
          \and
          M. Janssen\inst{7,1}
          \and
          D.~G. Nair\inst{8}
          \and
          T. Savolainen\inst{9,10,1}
          \and 
          E. Ros\inst{1}
          \and
          U Bach\inst{1}
          \and 
          J.~A. Hodgson\inst{11}
          \and
          M. Lisakov\inst{1,12}
          \and
          N.~R. MacDonald\inst{13}
          \and
          J.~A. Zensus\inst{1}
          }

   \institute{Max-Planck-Institut für Radioastronomie, Auf dem Hügel 69, D-53121 Bonn, Germany
            \and
             Theoretical Division, Los Alamos National Laboratory, Los Alamos, NM 87545, USA
            \and
            Instituto de Astrofísica de Andalucía-CSIC, Glorieta de la Astronomía s/n, E-18008 Granada, Spain
            \and
            NASA Marshall Space Flight Center, Huntsville, AL 35812, USA
            \and
            Institute of Astrophysics, Foundation for Research and Technology, Hellas, Heraklion 7110, Greece
            \and
            Department of Physics, Ulsan National Institute of Science and Technology (UNIST), 50 UNIST-gil, Eonyang-eup, Ulju-gun, Ulsan 44919, Republic of Korea
            \and
            Department of Astrophysics, Institute for Mathematics, Astrophysics and Particle Physics (IMAPP), Radboud University, PO Box 9010, 6500 GL Nijmegen, The Netherlands
            \and 
            Astronomy Department, Universidad de Concepción, Casilla 160-C, Concepción, Chile
            \and
            Aalto University Department of Electronics and Nanoengineering, PL 15500, 00076 Aalto, Finland
            \and 
            Aalto University Metsähovi Radio Observatory, Metsähovintie 114, 02540 Kylmälä, Finland
            \and
            Dept. of Physics \& Astronomy, Sejong University, Guangjin-gu, Seoul 05006, Republic of Korea
            \and
            Instituto de Física, Pontificia Universidad Católica de Valparaíso, Casilla 4059, Valparaíso, Chile
            \and
            Department of Physics and Astronomy, The University of Mississippi, University, Mississippi 38677, USA
             }

   \abstract{
AGN-launched jets are a crucial element in the study of super-massive black holes (SMBH) and their closest surroundings. 
The formation of such jets, whether they are launched by magnetic field lines anchored to the accretion disc or directly connected to the black hole’s (BH) ergosphere, is the subject of ongoing, extensive research.

3C\,84, the compact radio source in the central galaxy NGC\,1275 of the Perseus super-cluster, is a prime laboratory for testing such jet launching scenarios, as well as studying the innermost, sub-parsec AGN structure and jet origin. 
Very long baseline interferometry (VLBI) offers a unique view into the physical processes in action, in the immediate vicinity of BHs, unparalleled by other observational techniques. 
With VLBI at short wavelengths particular high angular resolutions are obtained.

Utilising such cm and mm-VLBI observations of 3C\,84 with the European VLBI Network and the Event Horizon Telescope, we study the magnetic field strength and associated accretion flow around its central SMBH. 
This is possible, as higher frequency VLBI measurements are capable of peering through the accretion flow surrounding the central engine of 3C\,84, which is known to block the line of sight to the sub-parsec counter-jet via free-free absorption. 
Furthermore, we study the magnetic field’s signature in the core region, as manifested in polarised light. 
As part of this analysis we compare our observations to relativistic magneto-hydrodynamic simulations. 
Finally, we investigate the effect of instabilities on the shape of the jet's parsec-scale funnel and try to connect them to its historical evolution.
   }

   \maketitle
%

\section{Open jet formation questions}
Astrophysical jets launched by active galactic nuclei (AGN) are powerful, highly collimated outflows of matter that play a crucial role in shaping the evolution of galaxies and their surrounding environments. 
They may influence star formation, regulate galaxy growth, and are thought to be able to transport energy over vast cosmic distances, impacting the interstellar, intergalactic, and intracluster media, thus providing insights into the most extreme physical conditions in the universe.
The exact mechanisms behind the formation of such astrophysical jets remain elusive.
Two leading models, presented in \cite{Blandford82} (BP) and \cite{Blandford77} (BZ), offer alternative theories: the former model suggests jets are driven by magnetic fields interacting with accreting material, while the latter model attributes their formation to the extraction of rotational energy from a spinning black hole.

\section{A laboratory in space}
3C\,84 (NGC\,1275), the brightest radio source in the Perseus cluster, is a prime laboratory for investigating jet formation due to its proximity and the unique properties of its AGN, thus being a frequent target of monitoring campaigns (e.g. presented in \cite{Paraschos22}).
Its jets are relatively young and evolving on observable timescales, facilitating the study of their structure and behaviour.
To distinguish between the models describing the launching and collimation of these jets, we rely on very long baseline interferometry (VLBI). 
VLBI provides the high resolution needed to observe the fine-scale structures in the jet close to the black hole, resolving features that would otherwise be unresolvable.
Interestingly, 3C\,84 exhibits an elongated structure in the core region, perpendicularly oriented to the bulk jet flow, as revealed by space-VLBI in \cite{Giovannini18}.
Attempts have been made to explore this region both spectrally (see, e.g. work by \cite{Paraschos21}) and via magnetic field signatures embedded in polarised light (see, e.g. work by \cite{Kim19}).
In the section below we discuss further insights about jet launching, that we gained by using polarised light observations of 3C\,84.

\section{EVN and EHT synergy}

Combining VLBI observations both at centimetre and millimetre wavelengths offers the unique opportunity to study the jet both in the ultimate vicinity of the central engine as well as further downstream in the sub-parsec scales.
We achieved this by observing the 3C\,84 jet with the European VLBI Network (EVN) at 22\,GHz and the Event Horizon Telescope (EHT) at 230\,GHz.

\subsection{Insights from cm-VLBI and simulations}

\begin{figure*}
   \centering
   \vspace{250pt}
   \includegraphics{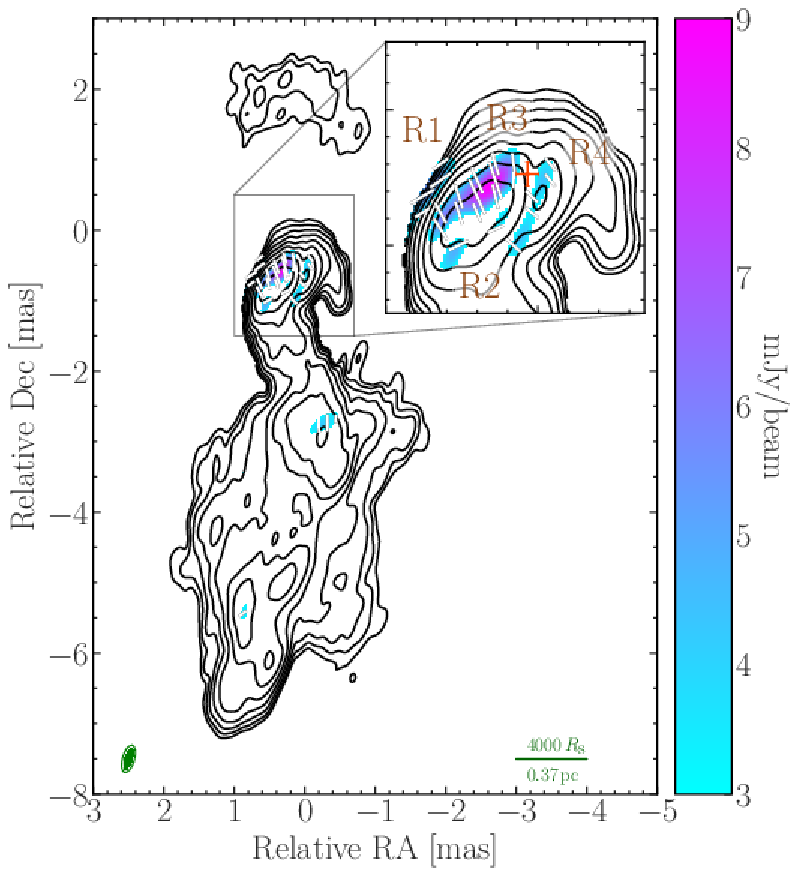}
   \includegraphics{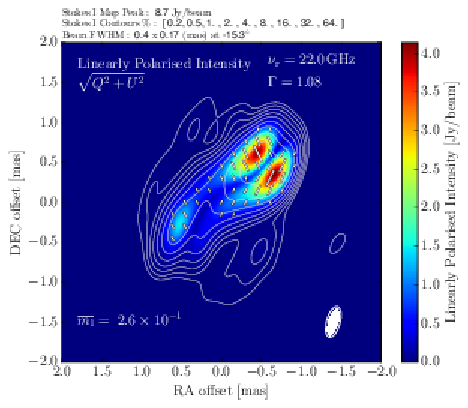}
   \caption{Observations and simulations of the 3C\,84 small-scale jet.
            Left: EVN image of Stokes I (black contours), polarised intensity (colour scale), and EVPA (white bars) of 3C\,84 at 22\,GHz.
            Right: Results from RMHD simulations showing a synthetic intensity map of a ray-traced hybrid fluid-particle jet, moving from north-west to south-east; Stokes I, P and EVPAs are shown similarly to the observations (as published in \cite{Paraschos24b}).}
    \label{fig:EVN}
    \end{figure*}

Our cm-VLBI observations revealed the presence of a double rail structure (also seen in \cite{Giovannini18}) in Stokes I but also in Stokes P deep in the core region (see Fig.~\ref{fig:EVN} left panel).
Furthermore, an inner spine is also detected in linear polarisation, at a significance level of $4\,\sigma$.
Interestingly, this jet stratification is accompanied by distinct electric vector position angles (EVPAs) along the two regions.
Specifically, the EVPAs are parallel at the spine to the bulk jet flow and perpendicular to it at the sheath.
A fourth region of interest is identified at the north east part of the core, spatially coincident with the area, where the jet is turning downstream.
The morphology of the EVPAs there is suggestive of a shock or jet shear.

We also compared our observations with state-of-the-art relativistic magneto-hydrodynamic (RMHD) simulations (see also the work by \cite{Kramer21, Kramer24}), which revealed that the observed EVPA pattern is consistent with a helical or toroidal magnetic field configuration (see Fig.~\ref{fig:EVN} right panel).
A poloidal field on the other hand did not reproduce the observed geometry.
Our RMHD simulations showed that the EVPAs also follow the bulk jet flow at the spine and are orthogonal to it at the limbs when the magnetic field is helical or toroidal. 
These patterns is also consistent with spine-sheath geometry.
A more detailed overview is presented in \cite{Paraschos24b}, on which this subsection is also based on.


%

\subsection{Insights from mm-VLBI}

\begin{figure*}
   \centering
   \vspace{200pt}
   \includegraphics{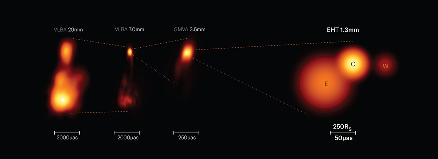}
   \caption{Jet morphology of 3C\,84 shown in total intensity at different wavelengths. 
            From left to right, we show the 15, 43, 86 (images), and 228 GHz (model) measurements (as published in \cite{Paraschos24a}).}
    \label{fig:EHT}
    \end{figure*}

\begin{figure}
   \centering
   \vspace{200pt}
   \includegraphics{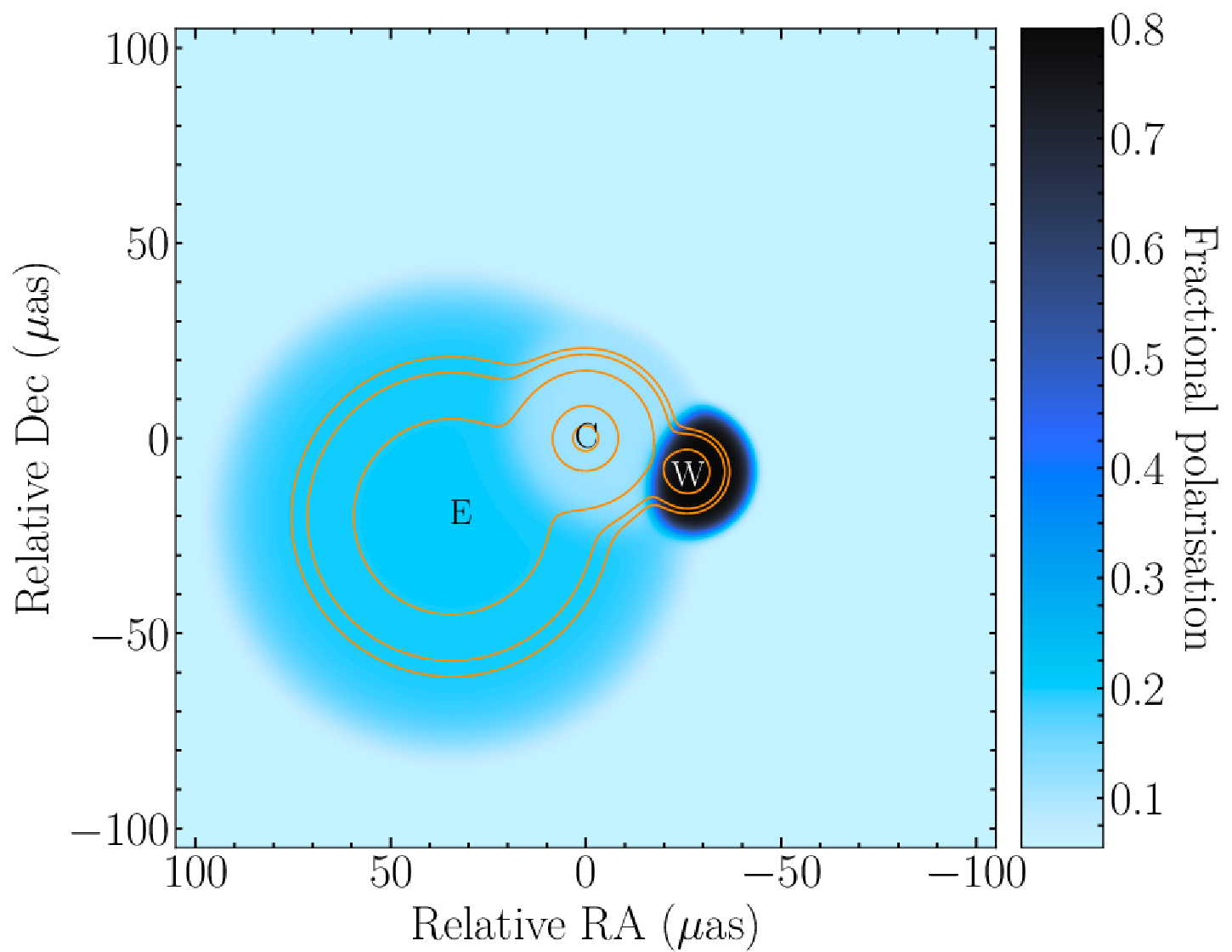}
   \caption{Fractional polarisation of the EHT data. 
    Shown here is a representation of the best-fit model to the fractional polarisation data in the image plane (as published in \cite{Paraschos24a}).}
    \label{fig:Frapol}
   \end{figure}

In order to peer even deeper into the core region, we employed mm-VLBI observations of 3C\,84, taken quasi-simultaneously with longer wavelength observations (see Fig.~\ref{fig:EHT}).
At 1.3\,mm the core of 3C\,84 is best modelled by three components called W, C, and E.
As shown in Fig.~\ref{fig:Frapol} component C, identified as the core, appears unpolarised.
On the other hand, the two components E and W framing it, exhibit high amounts of fractional polarisation (20-80\%).
Thus, E and W could correspond to the footpoints of the emerging jet.

Additional information can be gathered by leveraging the lower frequency observations to calculate physical quantities such as the Faraday rotation measure, flux density, and mass accretion rate.
From these quantities we were able to derive that magnetic field $B\sim3-6$\,G and also to show a preference for the spin parameter $\alpha_*=1$ (corresponding to rapid rotation associated with an advection dominated accretion flow; ADAF), and the dimensionless magnetic flux $\phi\sim40-90$ (indicating a preference for a magnetically arrested disc; MAD).
Evidence for the BZ type of jet launching is also found (consistent with \cite{Paraschos23}).
A more detailed overview is presented in \cite{Paraschos24a}, on which this subsection is also based on.


%

\subsection{A proposed toy model}

\begin{figure}
   \centering
   \vspace{200pt}
   \includegraphics{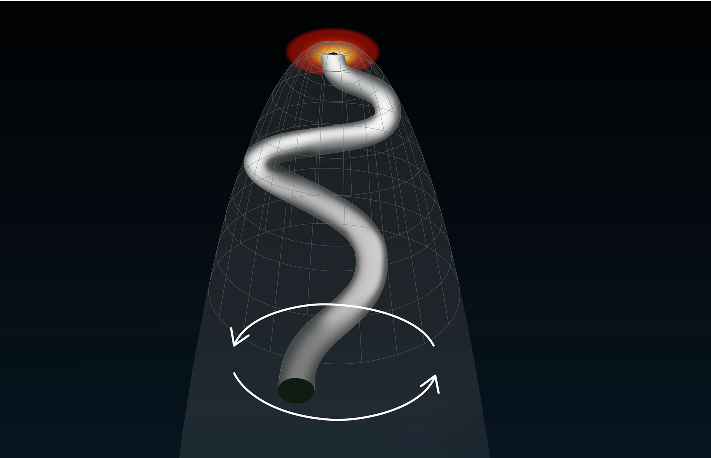}
   \caption{Sketch of a filamentary flux-tube structure rotating inside a surrounding sheath.}
    \label{fig:Tube}
    \end{figure}

A stratified combination of both a BP and BZ jet, consisting of a fast, rotating inner filament and an outer, slower moving sheath could be a potential explanation of the observed phenomenology (see Fig.~\ref{fig:Tube} for a visualisation).
This setup could produce the elongated core as a broad jet base associated with accretion disc jet launching, as well as the BZ type characteristics that our polarisation analysis revealed.
It would also explain naturally the jet stratification manifested in polarised light in the core of 3C\,84 in polarised light.
Finally, it could explain the observed moving jet components over the years (see, e.g. \cite{Paraschos22}) and the associated Doppler boosting as elements of the inner spine being aligned to our line of sight during parts of its rotation.

%

%

%

%
   
%

\section{Conclusions}
Our conclusions can be summarised as follows:
\begin{itemize}
    \item[$\bullet$] We found a highly coherent, strong magnetic field around the central SMBH of 3C\,84.
    \item[$\bullet$] The system indicates a preference for an advection-dominated accretion flow (ADAF) in a magnetically arrested (MAD) state around a rapidly rotating ($\alpha_*\sim1$) SMBH.
    \item[$\bullet$] On a larger scale, we found that the polarised emission in the core region traces the brightened limbs.
    \item[$\bullet$] The EVPA orientation suggests a toroidal or helical magnetic field configuration consistent with a spine/sheath geometry.
\end{itemize}

Our work presented here elucidates the excellent capability of combining cm and mm-VLBI to comprehensively study the jets of nearby AGN at multiple scales in both Stokes I and P.

\begin{acknowledgements}
The European VLBI Network (www.evlbi.org) is a joint facility of independent European, African, Asian, and North American radio astronomy institutes. Scientific results from data presented in this publication are derived from the following EVN project code(s): FILL HERE.
e-MERLIN is a National Facility operated by the University of Manchester at Jodrell Bank Observatory on behalf of STFC. 
The Very Long Baseline Array and the Green Bank Telescope is/are operated by The National Radio Astronomy Observatory, a facility of the National Science Foundation operated under cooperative agreement by Associated Universities, Inc.
This work made use of the Swinburne University of Technology software correlator, developed as part of the Australian Major National Research Facilities Programme and operated under licence.
This research has made use of data obtained using the Global Millimetre VLBI Array (GMVA), which consists of telescopes operated by the Max-Planck-Institut für Radioastronomie (MPIfR), IRAM, Onsala, Metsähovi Radio Observatory, Yebes, the Korean VLBI Network, the Greenland Telescope, the Green Bank Observatory (GBT) and the Very Long Baseline Array (VLBA). 
ALMA is a partnership of ESO (representing its member states), NSF (USA) and NINS (Japan), together with NRC (Canada), MOST and ASIAA (Taiwan) and KASI (Republic of Korea), in cooperation with the Republic of Chile. The Joint ALMA Observatory is operated by ESO, AUI/NRAO and NAOJ. 
The GLT is part of the ALMA–Taiwan project and is supported in part by the Academia Sinica (AS) and the Ministry of Science and Technology of Taiwan.
his research is based in part on observations obtained with the 100-m telescope of the MPIfR at Effelsberg, observations carried out at the IRM 30-m telescope operated by IRAM, which is supported by INSU/CNRS (France), MPG (Germany) and IGN (Spain), observations obtained with the Yebes 40-m radio telescope at the Yebes Observatory, which is operated by the Spanish Geographic Institute (IGN, Ministerio de Transportes, Movilidad y Agenda Urbana), and observations supported by the Green Bank Observatory, which is a main facility funded by the NSF operated by the Associated Universities. We acknowledge support from the Onsala Space Observatory national infrastructure for providing facilities and observational support. The Onsala Space Observatory receives funding from the Swedish Research Council through grant no. 2017-00648. This publication makes use of data obtained at the Metsähovi Radio Observatory, operated by Aalto University. 
This research makes use of VLBA data from the VLBA-BU Blazar Monitoring Program (BEAM-ME and VLBA-BU-BLAZAR; http://www.bu.edu/blazars/BEAM-ME.html), funded by NASA through the Fermi Guest Investigator Program.
This research has made use of data from the MOJAVE database that is maintained by the MOJAVE team (Lister et al. 2018).
This publication acknowledges project M2FINDERS, which is funded by the European Research Council (ERC) under the European Union’s Horizon 2020 research and innovation programme (grant agreement no. 101018682).  
D.G.N. acknowledges funding from Conicyt through Fondecyt Postdoctorado (project code 3220195).
\end{acknowledgements}


\begin{thebibliography}{}

  \bibitem[Blandford \& Znajek 1977]{Blandford77} {Blandford} R.~D. and {Znajek} R.~L. 1977, in MNRAS, {179}, {433-456}
  \bibitem[Blandford \& Payne 1982]{Blandford82} {Blandford} R.~D. and {Payne} D.~G. 1982, in MNRAS, {199}, {883-903}
  \bibitem[Giovannini et al. 2018]{Giovannini18} {Giovannini} G., {Savolainen} T., {Orienti} M. et al. 2018, in NatAst, {2}, {472-477}
  \bibitem[Kim et al. 2019]{Kim19} Kim J. -Y., Krichbaum T.~P., Marscher A.~P. et al. 2019, in A\&A, 622A, 196K
  \bibitem[{Paraschos} et al. 2021]{Paraschos21} {Paraschos} G.~F., {Kim} J. -Y., {Krichbaum} T.~P. et al. 2021, in A\&A, 650L, 18P
  \bibitem[Kramer \& MacDonald 2021]{Kramer21} {Kramer} J.~A. and {MacDonald} N.~R. in A\&A, 2021, 656A, 143K
  \bibitem[Paraschos et al. 2022]{Paraschos22} Paraschos G. F., Krichbaum T. P., Kim J. -Y. et al. in A\&A, 665A, 1P
  \bibitem[Paraschos et al. 2023]{Paraschos23} Paraschos G.~F., Mpisketzis V., Kim J.-Y. et al. 2023, in A\&A, 669A, 32P
  \bibitem[Paraschos et al. 2024a]{Paraschos24a} Paraschos G.~F., Kim J.-Y. Wielgus M. et al. 2024, in A\&A, 682L, 3P
  \bibitem[Paraschos et al. 2024b]{Paraschos24b} Paraschos G.~F., Debbrecht L.~C., Kramer J.~A. et al. 2024, in A\&A, 686L, 5P
  \bibitem[Kramer et al. 2024]{Kramer24} {Kramer} J.~A. and {MacDonald} N.~R., Paraschos G.~F. et al. in A\&A, 2024, arXiv:2409.05256



\end{thebibliography}
\end{document}